\documentclass[twocolumn,twocolappendix]{aastex7}

\usepackage[utf8]{inputenc}
\usepackage{tipa}
\usepackage{combelow}
\usepackage{savesym}
\savesymbol{tablenum}
\usepackage{siunitx}
\restoresymbol{SIX}{tablenum}
\usepackage{breqn}
\usepackage{tabularx}
\usepackage{placeins}
\usepackage{paralist}

\newcommand\kh{\textdoublebarpipe K\cb{\'{a}}g\'{a}ra-!H\~{a}unu}
\newcommand\kagara{\textdoublebarpipe K\cb{\'{a}}g\'{a}ra}
\newcommand\haunu{\textdoublebarpipe !H\~{a}unu}
\newcommand{\sssb}[3]{#1 #2$_{#3}$}

\begin{document}

\title{Trans-Neptunian Binary Mutual Events in the 2020s and 2030s}

\author[orcid=0000-0002-1788-870X,sname='Proudfoot']{Benjamin Proudfoot}
\affiliation{Florida Space Institute, University of Central Florida, 12354 Research Parkway, Orlando, FL 32826, USA}
\email[show]{benp175@gmail.com}

\author[orcid=0000-0002-8296-6540, sname='Grundy']{Will Grundy} 
\affiliation{Lowell Observatory, 1400 W Mars Hill Rd, Flagstaff, AZ 86001, USA}
\affiliation{Northern Arizona University, Department of Astronomy \& Planetary Science, PO Box 6010, Flagstaff, AZ 86011, USA}
\email{}

\author[orcid=0000-0003-1080-9770, sname='Ragozzine']{Darin Ragozzine} 
\affiliation{Brigham Young University Department of Physics \& Astronomy, N283 ESC, Brigham Young University, Provo, UT 84602, USA}
\email{}

\begin{abstract}

Mutual events of trans-Neptunian binaries (TNBs) provide rare opportunities to measure the physical and orbital properties of small bodies in the outer solar system. However, successful observations of these events have been limited by uncertain predictions. Here, we present probabilistic predictions of TNB mutual events occurring through the 2030s, using high-precision non-Keplerian orbit solutions from the \textit{Beyond Point Masses} project combined with a Bayesian framework that propagates orbital and size uncertainties. Our methods generate distributions of event timing, duration, depth, and probability of occurrence, enabling direct assessment of observability. We provide predictions for five systems with ongoing or imminent mutual event seasons, including (38628) Huya, (58534) Logos-Zoe, (148780) Altjira, (469705) \kh, and (524366) \sssb{2001}{XR}{254}. 
Preparing for upcoming events with long-baseline light curve monitoring is vital, as events may be difficult to distinguish from a regular rotational light curve. Rapid dissemination of event detections will benefit the entire community, allowing predictions to be updated, ensuring that these rare mutual event opportunities can be fully exploited. 

\end{abstract}


\section{Introduction} 
\label{sec:intro}

\begin{figure*}[t]
    \centering
    \includegraphics[width=\textwidth]{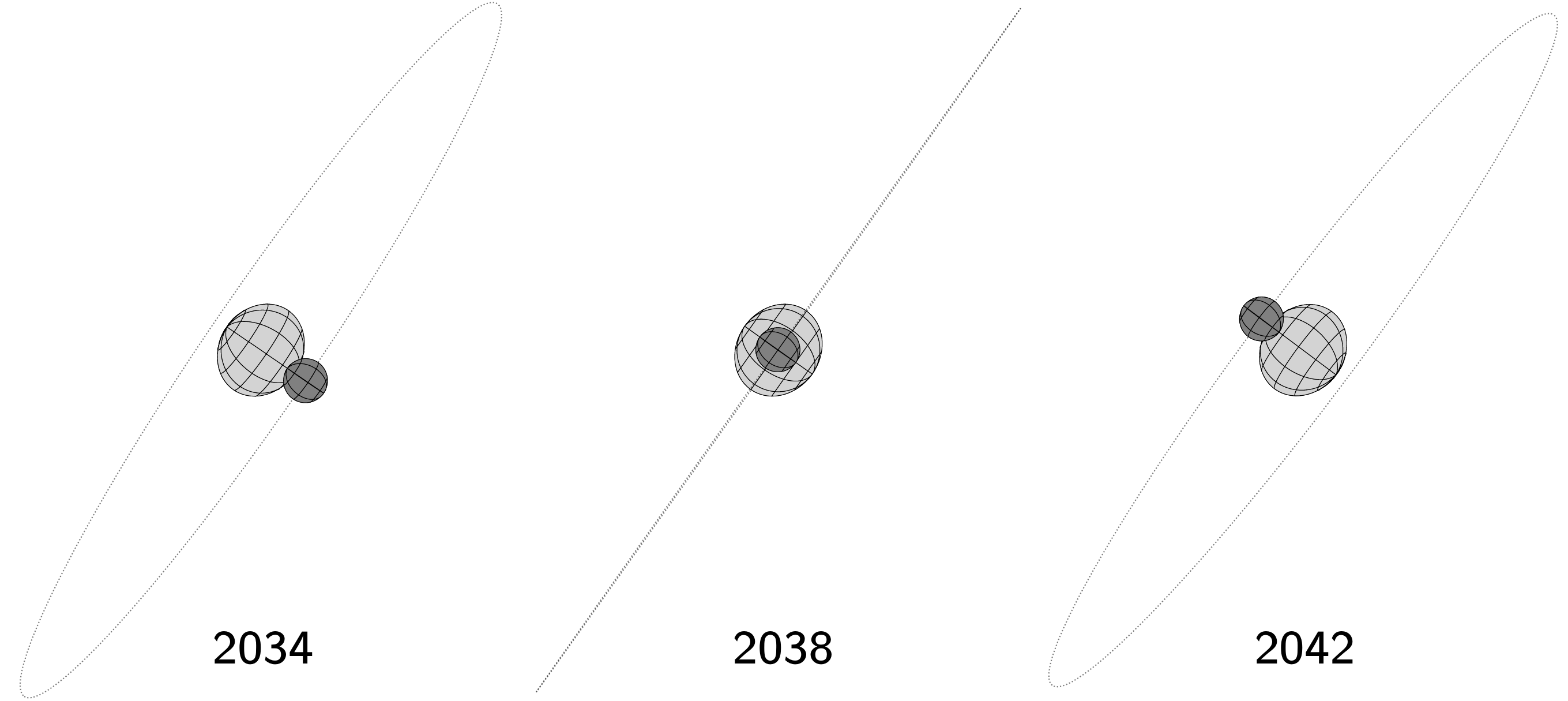}
    \caption{Mutual events of Huya throughout its 2033-2043 mutual event season. Huya's pole orientation \added{is} assumed to be the same as the satellite's orbit plane \citep{rommel2025huya}. This schematic does not show the effects of shadowing, which can modestly increase the depth of superior mutual events where the secondary is in the foreground.}
    \label{fig:huya}
\end{figure*}

Binaries are common in the trans-Neptunian region with more than 100 known binaries \citep[][]{noll2020trans}. Trans-Neptunian binaries (TNBs) are particularly important for study, as their mutual orbits provide various tools to further probe the physical characteristics of small trans-Neptunian objects (TNOs), allowing for measurements of mass, density, and tidal history of a given system \citep[e.g.][]{grundy2019mutual}. Twice throughout a TNB's centuries-long heliocentric orbit, the plane of its mutual orbit will closely align with our view from the inner solar system, creating a series of mutual eclipses and occultations between the binary components (see Figure \ref{fig:huya}). Observations of these events become a powerful tool for characterizing the individual components of a binary system \citep[][]{noll2020trans}. These events, referred to as mutual events, allow observers to measure the size \citep{dunbar1986modeling}, shape \citep{pinilla2022detection}, albedo \citep{buie1992albedo}, surface features \citep{1999AJ....117.1063Y}, thermal properties of each component \citep{mueller2010eclipsing}, and compositional differences \citep{buie1987water,wong2025evidence}.

Although used with great success to characterize asteroids in a variety of populations \citep[e.g.][]{mottola2000mutual,berthier2020physical,scheirich2022preimpact}, mutual events have not received as much attention among the TNBs. Early on, Pluto and Charon were closely characterized during their 1985-1990 mutual event season \citep{dunbar1986modeling,young1994new}, enabling precise measurements of size, which was later confirmed during the \textit{New Horizons} flyby \added{\citep{Stern2015,Nimmo2017}}. Mutual events were also used to create a primitive surface albedo map of Pluto \citep[][]{buie1992albedo,1999AJ....117.1063Y}. In addition to Pluto, the Sila-Nunam system, a near-equal sized, tidally evolved binary was characterized using a series of mutual events \citep{grundy2012mutual,benecchi2014ut}, but full analysis of those events has not yet been published. Other than these two examples, little further success has been had in observing TNB mutual events. Attempts to observe mutual events of Haumea-Namaka and Manwe-Thorondor were hindered by uncertain predictions which prevented successful detection and interpretation of those events \citep[e.g.][]{2009AJ....137.4766R,grundy2014orbit,rabinowitz2019complex,proudfoot2024beyond}. 

As mutual events provide an unrivaled opportunity to improve our knowledge of the physical characteristics of TNBs, in this work, we focus on providing accurate predictions of mutual events through the 2030s. 
To provide the best possible predictions, we use a three-fold approach: (1) new Hubble Space Telescope (HST) observations to refine TNB mutual orbits; (2) non-Keplerian ephemerides from the \textit{Beyond Point Masses} project \citep[e.g.][]{ragozzine2024beyond,nelsen2025beyond}; (3) a Bayesian framework to produce probabilistic event predictions and uncertainties.

\begin{figure*}
    \centering
    \includegraphics[width=\textwidth]{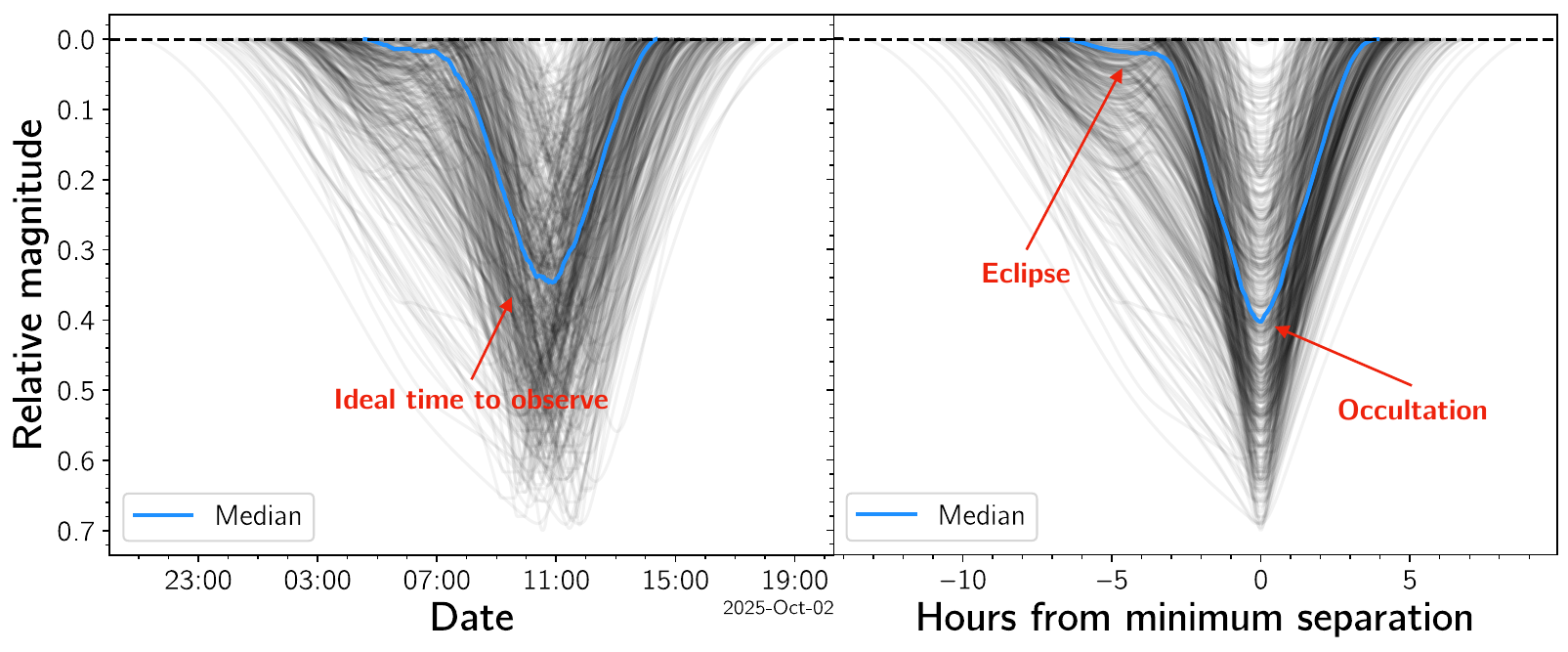}
    \caption{Our prediction of the light curves of Altjira's 2025-10-02 mutual event. Faint gray lines show the model light curves of posterior draws which show a mutual event (96\% of the 500). Blue lines show the median light curve of the ensemble of posterior draws (including those which produce no mutual event). On the left, light curves are displayed as a function of UTC date/time, where the median light curve corresponds to the expected magnitude drop at any given UTC time. Where the median light curve is non-zero are times at which observations are more likely than not (i.e. $p>0.5$) to catch a mutual event. On the right, model light curves are indexed to the time of minimum on-sky separation, where the median light curve corresponds to the typical light curve morphology. We note the median light curve does not necessarily represent a self-consistent mutual event light curve, but rather a statistical average incorporating uncertainties/differences in event timing, depth, duration, and morphology.}
    \label{fig:lc_good}
\end{figure*}

\section{Predicting Mutual Events}
\label{sec:methods}
All of our mutual event predictions use the \added{orbit solutions} provided by the \textit{Beyond Point Masses} project \citep{ragozzine2024beyond,proudfoot2024bpm2}. \added{These orbit solutions are combined with heliocentric ephemerides of the system, given by JPL Horizons, to provide precise ephemerides of each binary component. In this work, ephemerides} have been updated with the use of HST imaging from a variety of recent programs \citep[e.g.,][see also Appendix \ref{app:orbits}]{nelsen2025beyond}, in addition to detections from occultations of Huya and its satellite \citep{rommel2025huya}. In Appendix \ref{app:orbits}, we provide further details on new HST observations and the orbit solutions we use to predict mutual events. These orbit fits generate posterior solutions, made up of $\sim$millions of independent, statistical samples of each system's mutual orbit. These samples are useful for providing probabilistic mutual event predictions. 

With orbit posteriors of TNBs (soon to be) undergoing mutual events, we can predict \added{those} events using the method outlined in \cite{dunbar1986modeling} and \citet{brozovic2024orbit}. This treats each TNB component as a flat disk with a Lambertian surface (i.e. not accounting for the center-to-limb brightness distribution). With this model the normalized flux change during a mutual event is calculated as:

\begin{align}
\label{eqn:model}
    F = \frac{\kappa_2}{\kappa_1} \frac{\Delta A}{\pi (R_1^2 + \frac{\kappa_2}{\kappa_1}R_2^2)\cos^2(\alpha/2)}
\end{align}
\noindent where $\kappa$ is albedo, $\Delta A$ is the three-circle overlap area, $R$ is the component radius, and $\alpha$ is the phase angle. Subscripts refer to the primary ($1$) and secondary ($2$). The three-circle overlap area is the total area of the background object obscured by occultation/eclipsing by the foreground object. The $\cos^2(\alpha/2)$ term accounts for the gibbousness of the TNBs caused by non-zero phase angles. However, in practice, this term is negligible for the TNBs discussed in this work as $\alpha \lesssim2\degr$, which corresponds to a $\sim$0.05\% decrease in the illuminated area. Likewise, throughout this paper, we assume equal albedos of either component to simplify our model. Although this simplified photometric model is not fully accurate, it provides a firm foundation on which mutual event predictions can be made. 

Using the MCMC chains provided by the orbit fits in Appendix \ref{app:orbits}, we create an ephemeris \added{for} each TNB from 2025-2040, with a timestep of 2 hours \citep[similar to][]{proudfoot2024beyond,nelsen2025beyond}. Then, interpolating through this ephemeris, we identify candidate mutual events based on close-approaches between binary components (with separation $<$few primary radii). \added{Next,} we explicitly integrate a random posterior sample to the candidate epoch and create a new ephemeris \added{for} the system (based on a single posterior draw) in the few days around the possible mutual event. Using probabilistically drawn radii of each component, we then use our photometric model (Equation \ref{eqn:model}) to identify the photometric deviation (i.e. magnitude drop), times of first and last contact, time of minimum light, duration, and minimum separation during the mutual event. We repeat this process with 500 random posterior samples, allowing us to provide probabilistic predictions of the characteristics of a given mutual event, accounting for both orbit and size uncertainties. It also allows us to estimate the probability that the event will occur, as not all of the orbit/size posterior will necessarily produce a mutual event. We then repeat this process for each candidate mutual event identified. 

\begin{figure*}
    \centering
    \includegraphics[width=\textwidth]{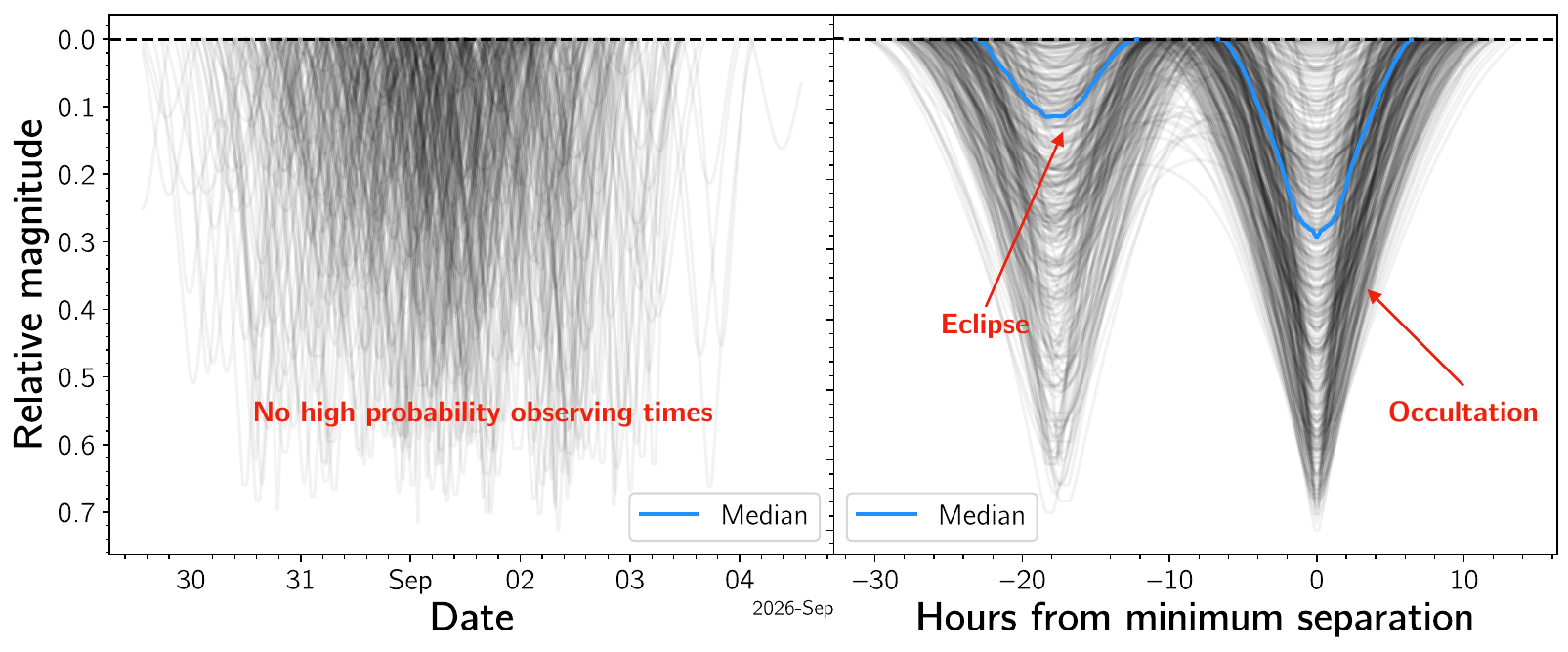}
    \caption{The predicted light curves of Altjira's 2026-09-01 mutual event, in the same style as Figure \ref{fig:lc_good}, though the range of potential mutual event times is much larger (several days compared to hours in Figure \ref{fig:lc_good}). Notably, in the left panel, there are no times at which observations are more likely than not (i.e. $p>0.5$) to catch an ongoing mutual event, despite $\sim97\%$ of statistical samples producing an observable mutual event. \added{In the right panel, when aligning the events to sample's time of minimum separation, a distinct double-peaked mutual event is apparent. This occurs as the viewing geometry provides significant separation between the foreground component and its shadow.}}
    \label{fig:lc_bad}
\end{figure*}

This method naturally allows us to create model light curves of each mutual event based on the radii and randomly drawn posterior sample. The ensemble of light curves from different posterior draws provide a rough prediction for the expected event light curve. We provide an example of this in the left panel of Figure \ref{fig:lc_good}. Since some events have large timing uncertainties, producing a messy cloud of event predictions, we also provide a predicted light curve where the timing is indexed to the moment of closest approach. This provides a better illustration of the morphology of the event, as seen in the right panel of Figure \ref{fig:lc_good}. This is especially relevant for events where the absolute timing uncertainties remain large. We show an example of this in Figure \ref{fig:lc_bad}.

\section{Upcoming Mutual Events}
\label{sec:predictions}

In this section, we describe upcoming and ongoing mutual event seasons \added{for} 5 different TNB systems. An overview of our results in \added{shown} in Table \ref{tab:overview}, and the full set of predictions are provided in Appendix \ref{app:tables}.

\begin{deluxetable*}{lcccccc}
\tabletypesize{\footnotesize}
\tablewidth{\textwidth}
\tablecaption{Mutual Events of TNBs}
\tablehead{
Name/Designation & Approx. $V$ & Event Season & $N$ & Timing Unc. & Typical Duration & Typical Depth \\
 & (mag) & & & (hours) & (hours) & (mag)
}
\startdata
(38628) Huya & 19.9    & 2033-2040+ & 1000+ & $<$1    & $<$5    & $\sim$0.25 \\
(58534) Logos-Zoe & 23.7   & 2027-2029  & 6                        & 11/45+ & 10/18+ & $\sim$0.5 \\
(148780) Altjira & 23.0 & 2025-2030  & 25                       & 1/20+ & 10/20 & $\sim$0.5  \\
(469705) \kh & 22.5 & 2025-2038  & 47                       & 7/12  & 8/40+ & $\sim$0.45 \\
(524366) \sssb{2001}{XR}{254} & 22.4     & 2031-2040+ & 46                       & 6/22  & 10/20 & $\sim$0.3  \\
\enddata
\tablecomments{Overview of TNB mutual events from the present through the 2030s. $N$ measures the number of events (with $P>0.05$) before 2040. Columns with two values show the uncertainty and duration of inferior and superior events.}
\label{tab:overview}
\end{deluxetable*}

\subsection{(38628) Huya}
Huya and its satellite have recently been explored in detail by occultations \citep{santos2022physical,rommel2025huya}. Based on 20+ years of astrometry of the satellite combined with several detections of the satellite in stellar occultations, \citet{rommel2025huya} was able to determine the mutual orbit of the Huya system \added{with $<1$ second precision on the mutual orbit period}. With this orbit fit, we predict that reasonably likely ($p>0.5$) grazing events will start in $\sim$2033. Grazing events will continue until $\sim$2037, at which point they become total occultations/eclipses with depth of $\sim0.25$ mag. Total events will continue until $\sim$2041, with grazing events until $\sim$2043. We show a schematic of the mutual event season progression in Figure \ref{fig:huya}. In total, we predict $>$2000 events throughout the mutual event season. Current timing uncertainties are $\lesssim1$ hour \added{throughout the 2030s.}

Huya is an ideal target for in-depth mutual event characterization as it is one of the brighter known TNOs. At $V\approx19.8$, mutual events should be observable with medium-sized telescopes. In addition, since Huya's light curve has been shown to be mostly flat, interpreting mutual events is \added{simpler than for system's with complex, high-amplitude light curves}. Huya rotates relatively rapidly \added{\citep[$6.725\pm0.006$ hr][]{santos2022physical}}, which allows successive mutual events to explore albedo features at all longitudes of Huya's surface. In comparison, since Pluto and Charon were tidally locked, mutual events could only probe the sub-Charon/sub-Pluto hemispheres of Pluto/Charon \citep{buie1992albedo,1999AJ....117.1063Y}. A broad, international collaboration of mutual event observers can enable precise determinations of Huya's size, shape, albedo variegation, thermal properties, and photometric behavior. 


\subsection{(58534) Logos-Zoe}
Logos-Zoe, a cold classical binary, is a wide system with a long 309.5 day orbit. \citet{thirouin2025logos} studied the system in detail, finding that Logos is likely a close or contact binary with a high amplitude rotational light curve. Although this can induce non-Keplerian precession, Zoe's long-period orbit makes this precession extremely slow. The long period also significantly complicates the mutual event season. Over its whole mutual event season, we predict only 6 events spread across two years. 

\added{One event (2028-03-19) is the most probable, although timing uncertainties are currently $>1$ day. This event should produce a $0.47\pm0.16$ mag drop. Across all 6 events, timing uncertainties are typically $\gtrsim$ 0.5 day, significantly reduced from predictions made prior to our new HST observations \citep[$\sim1$ week,][]{thirouin2025logos}. Events are long, lasting between 0.5-1 days. With only 6 events,} there are limited opportunities to characterize \added{this} system, so observing early events is of paramount importance. \added{Additional HST observations taken in 2026 will also help to improve the event predictions (Program 18124, PI: A. Thirouin)}.

With an interesting triple configuration, the shape and depth of the event light curve may significantly differ from our predictions. \citet{thirouin2025logos} explores this further, showing how event light curves are functions of the system architecture, rotational phase, and multiplicity. Crucially, that work shows that predictions are remarkably consistent, even when considering different architectures. This provides further confidence that our two-disk model produces trustworthy predictions. Future work incorporating shape models may provide some improvement over our predictions, but is unlikely to change the predictions substantially. On the other hand, more detailed models will be necessary for interpreting observations after event detection.

\subsection{(148780) Altjira}

The Altjira system was studied in detail in \citet{nelsen2025beyond}. There, they found that the mutual orbit of the unnamed secondary was significantly non-Keplerian. They propose that the system primary is possibly made up of a close binary---or possibly a contact binary---making this system an example of a hierarchical triple system, like the Lempo system \citep{benecchi201047171}. This complicates modeling of the mutual events, as the shape, separation, and orbital phase of the putative inner binary is unknown. It may be possible that each mutual event produces two distinct drops in light from each inner binary component. Further, it is plausible that the inner binary is also undergoing its own mutual event season, creating complex light curves of multiple mutual events occurring within a few hours. 

Since little is known about the putative inner binary, we proceed with modeling just a two-body system using our simplified photometric model. We use a non-Keplerian orbit posterior from \citet{nelsen2025beyond}, which accounts for precession caused by the putative inner binary. This allows timing predictions to be accurate, but likely does not capture the actual shape/duration of mutual events. Observers of events should plan to observe well-before/-after our listed first/last contact times to account for photometric modeling systematics (i.e. assuming a two-body model instead of a full three-body model). 

Events appear to be currently ongoing, with a total of 25 events (with $p>0.05$) from 2025-2030. Timing uncertainties on the shorter events near periapse---with durations of 5--10 hours---are $\lesssim1$ hour throughout the event season. Slower events have more uncertain timing ($\gtrsim20$ hours), partly due to their slower velocity, but also due to uncertainties in the secondary's precession rate. Typical depths of $\sim0.5$ mag are estimated using our two-disk model, but these could be substantially different based on the system architecture. 

Mutual events of the Altjira system act as a sensitive probe of the system architecture, which is otherwise impossible to currently probe. We expect an inner binary will cause a ``double'' event, where there appears to be two distinct drops in the light curve (or four drops depending on the separation between the shadowing and occultation, compare Figures \ref{fig:lc_good} and \ref{fig:lc_bad}). Alternatively, a contact-binary-like shape will exhibit a longer than expected drop, with a possibly asymmetric shape. Full analysis will require a dedicated forward modeling software like \texttt{Candela} \citep{thirouin2025logos}.

\subsection{(469705) \kh}
The mutual event season of \kh{} is currently ongoing, with visible events approximately once every two months. Ongoing events are deep, producing $\Delta m \approx 0.45$ mag. The mutual orbit is eccentric ($e\approx0.69$) and the orientation of the orbit aligns the longitude of periapsis along the line-of-sight, making mutual events occur near apo/pericenter. As such, superior events last $<$10 hours, while inferior events can last days. In addition, this orientation allows superior events to continue for far longer than inferior events, with probable ($p>0.5$) events ongoing until $\sim$2034.

\added{Timing uncertainties for the ongoing superior events are $<$7 hours through 2026, although inferior events have $12-14$ hour uncertainties until the end of inferior events in 2028.} With superior events lasting $\sim$7 hours, observations are likely to catch at least a partial event. A 10 hour observing block centered on minimum light has a $\sim$55\% chance of catching minimum light and a $\sim$80\% chance of catching a partial event. With even just a partial event detection, updated predictions can be far more precise. Long inferior events lasting days may prove difficult to detect, as the system will dim by $\lesssim0.03$ mag hour$^{-1}$\added{.}

\subsection{(524366) 2001 XR$_{254}$}
The mutual event season of 2001 XR$_{254}$ will begin in the early 2030s, although the exact start of events is uncertain. Our orbit fits have tentatively detected orbital precession (see Appendix \ref{app:orbits}), but with substantial uncertainties, the geometry of mutual events is not well constrained. This is reflected in the low event probabilities in Table \ref{tab:xr}. Even so, the timing uncertainty of inferior events is $\lesssim8$ hours through the 2030s. With inferior event durations of $\sim12$ hours, observations remain feasible. Probable events ($p>0.5$) begin in 2036, and continue through the decade into the 2040s. 

With such uncertainties in the event geometry, 2001 XR$_{254}$ is a good candidate for follow-up imaging. Even a single observation of the system in the late 2020s can collapse the uncertainties dramatically, while also narrowing timing uncertainties. Observations near maximum separation help to constrain the event geometry (by constraining the orbital plane), while observations during XR$_{254}$'s maximum on-sky velocity (due to mutual orbit, not heliocentric orbit) constrain the event timing. Similarly, observing a mutual event collapses timing uncertainties.

\subsection{Other systems}

In addition to the systems we explicitly examine above, a few other systems may have upcoming (or even ongoing) mutual events. The first is \sssb{2003}{UN}{284}, which is an ultra-wide TNB expected to reach the peak of its mutual event season in 2038. However, as an ultra-wide binary with an orbit period $P_{orb} = 8.4$ yr, mutual events are unlikely to occur. Even if the orbit phase and orientation happened to produce a mutual event, we expect the timing uncertainty on this event would be $\sim$weeks. 

Another possible set of upcoming TNB mutual events are those of \sssb{2002}{WC}{19}. Currently, this system has a mirror ambiguous orbit. The two degenerate orbit solutions predict mutual event seasons in either 2042 or 2085. Given its tight orbit and assuming the 2042 timeline, mutual events could start sometime in the late 2030s. Only one or two high resolution observations are capable of breaking the mirror ambiguity, so we encourage follow-up observations. 

Another promising candidate for current or future mutual events is \sssb{2013}{FY}{27}. Although currently the orbit of its satellite is unknown, it appears to be in an edge-on configuration \citep[][]{sheppard2018albedos}. Thus far, determining its orbit has proven difficult due to this configuration. The combination of orbit characterization (giving mass, density, etc.) and mutual events could provide a powerful probe into the physical characteristics of mid-sized TNOs, which are generally poorly explored. Similarly, Makemake's satellite is also on an edge-on orbit \citep{parker2016discovery,parker2018mass}. These systems are currently being examined in ongoing HST programs \citep[PID 18006 and 18133,][]{makemakehst,fy27hst}.

We leave predicting the mutual events of these systems to future work.

\subsection{(42355) Typhon-Echidna}
Previous studies observing the centaur binary Typhon-Echidna predicted mutual events observable around 2019--2026 based on a Keplerian orbit fit \citep{grundy200842355}. Our HST imaging in 2025 also targeted this system, to determine if mutual events were still ongoing. We found, however, that the mutual orbit has \textit{significantly} precessed---with nodal precession rates of $\sim4\degr$ yr$^{-1}$ around a pole $25\degr$ inclined to the mutual orbit plane---radically changing its mutual orbit plane in the two decades since its discovery. This highlights the importance of accounting for non-Keplerian effects in the predictions of mutual events. The results of our study of the Typhon-Echidna system will be explored more fully in a forthcoming publication.

\begin{figure}
    \centering
    \includegraphics[width=\columnwidth]{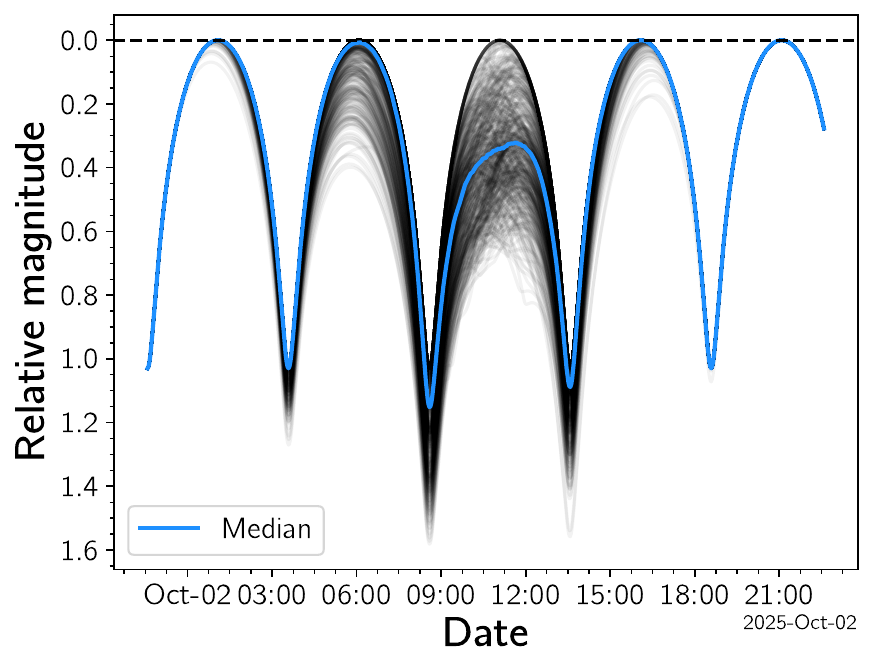}
    \caption{An example of the light curve of Altjira's 2025-10-02 mutual event if the system has a contact-binary-like light curve. In contrast with Figure \ref{fig:lc_good}, we display all modeled light curves---even if they do not produce a mutual event. For this date, 96\% of the 500 samples produce a mutual event.}
    \label{fig:withlc}
\end{figure}

\section{Discussion}
\label{sec:discussion}

Throughout this work, we discuss the probability of ``catching'' a mutual event by \added{taking observations} at a given time. Although this provides a good guide of when to observe, we note that this only accounts for a single snapshot in time. For an observing run that lasts many hours, there is a much larger probability of catching at least a partial event. For example, take the 2026-03-30 mutual event of \kh{} with timing uncertainty on minimum light of $\sim6.5$ hours and an event duration of $\sim7$ hours. If observed for 10 hours continuously centered on the predicted event mid-time, there is a $\sim56\%$ chance of observing minimum light, but a $\sim81\%$ chance of observing at least part of the mutual event. As a rough heuristic, an event is ``worth'' observing if the timing uncertainty is equal to or less than the event duration. All our event predictions roughly follow this, except Logos-Zoe. 

For each mutual event season, particular focus should be dedicated to early events. Detecting early events will allow for improved predictions---in both event timing and geometry---which will allow later observations to be more carefully planned. Even partial events can \textit{dramatically} improve predictions \citep{benecchi2014ut}.

Interpretation of mutual event light curves---or even understanding if a mutual event has been detected---can be difficult. Although seemingly obvious in Figure \ref{fig:lc_good} and \ref{fig:lc_bad}, any drop in brightness will be overlaid atop the system's combined light curve. If complex, the system light curve can substantially obscure mutual events. In Figure \ref{fig:withlc}, we show our 2025-10-02 Altjira event light curve (from Figure \ref{fig:lc_good}) atop a 1 mag amplitude contact-binary light curve. Further, binaries with near-equal sized objects can have complex light curves made up of two light curves superimposed upon one another. With no knowledge of the system light curve beforehand, a true observation of a mutual event could be mistaken for a non-detection, or vice versa. This is best prevented by monitoring of the system light curve before and after any events. Long-term monitoring in the years leading up to (or after) events must also play a role. See \citet{thirouin2025logos} for more detail.

We note that the Vera C. Rubin Observatory Legacy Survey of Space and Time (LSST) will make hundreds of photometric measurements of many TNOs \citep[e.g.,][]{kurlander2025predictions}. While these measurements will be sparse, they will be numerous, well-calibrated, and reasonably precise. Thus, it should be possible for most TNOs to derive lightcurves. For TNBs (known and unknown) where each component is reasonably detectable in single LSST exposures (e.g., brigther than about 23rd magnitude), it may be possible to detect two periods, especially if the two objects are occasionally resolved \citep[e.g.,][]{rommel2025spintrace}. Characterization of mutual events will be supported by LSST photometry which will provide useful lightcurve information (including phase curves and colors) that can be used to characterize the system lightcurve outside of mutual events. 

In the case of short-period binaries observed hundreds of times, like Huya, LSST should serendipitiously make some observations during a mutual event ($\sim$3\% of the time). TNOs that are not known to be binaries (or which are hierarchical binaries like proposed for Altjira) may have even closer orbital periods and thus higher likelihood ($\sim \frac{R}{a}$) for serendipitous mutual events. This is common in the asteroid belt where many binaries are regularly discovered through mutual events in photometry \citep{pravec2006photometric}. Over the course of its 10-year survey, identifying the frequency of mutual events in LSST TNO lightcurves could thus provide unique insights into the population of otherwise $\sim$undetectable close binaries. 

\section{Conclusions}
We present probabilistic predictions of mutual events for TNB systems expected to experience observable events through the 2030s. These predictions combine recent high-precision orbit fits with a Monte Carlo framework to quantify uncertainties in event timing, depth, and geometry. We find the following key results:

\begin{itemize}
    \item Probabilistic mutual event prediction provides several key benefits, including well determined timing/duration uncertainties and event probabilities, enabling robust observation planning.
    
    \item Mutual events are expected or ongoing in (at least) 5 TNB systems in the coming decades. These events provide numerous opportunities to characterize TNBs from a variety of populations. 

    \item Observing mutual events of some TNBs---like Altjira or Logos-Zoe---may reveal complex system architectures made up of more than two bodies. Mutual events provide a unique way to characterize and detect triple (or higher multiplicity) systems that are below the resolution of current (or even next generation) telescopes. 

    \item Broad community collaboration and preparation for mutual events is needed. Collaboration between observers at different longitudes can increase temporal coverage around the expected mutual event time. Preparation for mutual events should focus on understanding a TNB's light curve. Complex light curves can make interpreting---or even identifying---mutual events difficult, so light curve observation campaigns are vital to robust interpretation of mutual events. 

    \item Timely observations of early mutual events, even if only partial detections, collapse timing uncertainties on future events dramatically. Rapid, public dissemination of event detections will enable the community to maximize the scientific return of these one-in-a-century+ mutual events.
\end{itemize}

Mutual events in the coming decades allow a rare window into the physical and orbital properties of TNBs. Coordinated, well-timed observations will enable detailed characterization of size, shape, albedo, albedo variegation, and system architecture---offering key insights into binary formation and the evolutionary history of the outer solar system. With thoughtful planning and community coordination, the upcoming decades offer an opportunity to dramatically expand our understanding of TNBs.

\begin{acknowledgments}
\added{We thank Audrey Thirouin for discussions which helped to improve our manuscript. We also thank an anonymous reviewer whose comments substantially improved the clarity and readability of our manuscript.} We acknowledge the BYU Office of Research Computing for their dedication to providing computing resources without which this work would not have been possible. 

This research is based on observations made with the NASA/ESA Hubble Space Telescope obtained from the Space Telescope Science Institute, which is operated by the Association of Universities for Research in Astronomy, Inc., under NASA contract NAS 5–26555. These observations are associated with program 17707. Support for Program number 17707 was provided through a grant from the STScI under NASA contract NAS5-26555.

B.P. acknowledges the support of the University of Central Florida Preeminent Postdoctoral Program (P$^3$). 
\end{acknowledgments}

\begin{contribution}

B.P. led the overall analysis, writing of the paper, designing/scheduling of observations, and was the principal investigator of the HST program. 

W.G. analyzed all observations (both new and archival), contributed to interpretation, and provided editing support. 

D.R. provided access to computing resources, contributed to interpretation, and provided editing support.


\end{contribution}

\appendix
\restartappendixnumbering

\section{New Astrometry and Orbit Solutions}
\label{app:orbits}

\begin{deluxetable*}{ccccCCCC}
\tablecaption{New observations of systems near mutual event seasons}
\tablehead{
Name/Designation & Julian Date & Date & Telescope/Instrument & $\Delta\alpha\cos{\delta}$ & $\Delta\delta$ & $\sigma_{\alpha}$ & $\sigma_{\delta}$ \\
 & & & & \textrm{('')} & \textrm{('')} & \textrm{('')} & \textrm{('')}
}
\startdata
Logos-Zoe & 2460676.22927 & 2024-12-31 & HST/WFC3 & -0.02071 & -0.11942 & 0.00258 & 0.00142 \\
\nodata & 2460751.35659 & 2025-03-16 & HST/WFC3 & -0.01014 & +0.11703 & 0.00149 & 0.00165 \\
2001 XR254 & 2460622.82663 & 2024-11-08 & HST/WFC3 & -0.03419 & +0.00156 & 0.00368 & 0.00253 \\
\nodata & 2460651.13633 & 2024-12-06 & HST/WFC3 & -0.31810 & +0.22889 & 0.00065 & 0.00032 \\
\kh & 2460712.91737 & 2025-02-06 & HST/WFC3 & -0.15372 & +0.10348 & 0.00072 & 0.00068 \\
\nodata & 2460736.07585 & 2025-03-01 & HST/WFC3 & -0.14288 & +0.09419 & 0.00045 & 0.00039 \\
\enddata
\tablecomments{Astrometry is referenced to the primary system component based on the expected position from previous orbit fits. For one of the observations of Logos-Zoe, Zoe was brighter than Logos, likely due to the components' unusual light curves \citep{thirouin2025logos}. }
\label{tab:observations}
\end{deluxetable*}

Observations of four binaries---Logos-Zoe, \sssb{2001}{XR}{254}, \kh, and Typhon-Echidna---were acquired with HST as part of program 17707 (PI: Proudfoot) to improve predictions for ongoing or imminent mutual event seasons. Observations were taken with Wide Field Camera 3 (WFC3) using the broad F350LP filter to maximize SNR and therefore astrometric precision. Calibrated observations (using the standard WFC3 pipeline) were analyzed using standard PSF fitting methods using TinyTim model PSFs \citep{krist201120}. Our astrometry pipeline has been well-tested and has a long legacy of use for astrometry of TNBs \citep[e.g.,][]{grundy2019mutual}. We provide our new astrometric data in Table \ref{tab:observations}, except in the case of Typhon-Echidna, which will be discussed in a forthcoming publication. All new HST observations used in this work can be accessed at \dataset[DOI: 10.17909/jq61-jz73]{http://dx.doi.org/10.17909/jq61-jz73}.

Using both our new data, as well as published observations \citep{grundy2009mutual,grundy2011five,grundy2019mutual}, we refit the orbits of Logos-Zoe, \sssb{2001}{XR}{254}, and \kh with \texttt{MultiMoon}, a non-Keplerian orbit fitter designed for use with TNBs \citep{ragozzine2024beyond}. We performed orbit fits using a non-Keplerian framework, which allows us to constrain non-Keplerian precession in the mutual orbits, further allowing us to propagate those uncertainties to mutual event predictions. \texttt{MultiMoon} fits, in terms of parameterization, priors, likelihood functions, walker settings, burn-in, etc., were conducted similar to previous works \citep[see][]{proudfoot2024bpm2,nelsen2025beyond}. 

The goodness-of-fit for the Logos-Zoe and \sssb{2001}{XR}{254} orbit solutions was adequate ($\chi^2$ per degree of freedom $<1$), but was slightly elevated for \kh{} ($\chi^2_{\rm pdf}\sim2$). This could indicate some issues with data quality from heterogeneous sources (Keck, Gemini, HST/WFPC2, HST/WFC3), indeed almost one-third of the $\chi^2$ is due to a single data point near the middle of the observational baseline. Alternatively, higher-order non-Keplerian effects that our model does not capture may be present. Despite this, typical residuals are still $\sim7$ mas and the uncertainty in the separation between \kagara{} and \haunu{} is $<10$ mas. The majority of this uncertainty is accounted for in the event timing uncertainty. As such, we expect that our mutual event predictions will not be overly affected.

We note that for \sssb{2001}{XR}{254}, we find some indication that non-Keplerian effects are necessary to adequately describe the orbital motion. A Keplerian orbit solution provided a $\chi^2=24.6$ with 12 degrees-of-freedom, while the non-Keplerian provided a $\chi^2=8.2$ with 9 degrees-of-freedom. Although we can only reject the Keplerian fit at $2.4\sigma$ confidence, this again confirms the necessity of including non-Keplerian effects when predicting mutual events.

In Table \ref{tab:orbits}, we show the orbit solutions used for our probabilistic mutual event predictions. For binaries not discussed above, mutual orbit solutions are taken directly from our previous work (as referenced in the table).

\begin{deluxetable*}{lcccccc}
\tabletypesize{\footnotesize}
\tablecaption{Orbit solutions used for predicting mutual events}
\tablewidth{0pt}
\tablehead{
\multicolumn{2}{c}{Parameter} & \multicolumn{5}{c}{Posterior Distribution}
}
\startdata
    &               & \textbf{Huya}             & \textbf{Logos-Zoe}        & \textbf{Altjira}          &\textbf{\kh{}}             & \textbf{\sssb{2001}{XR}{254}} \\
    \\
Mass, Primary ($10^{18}$ kg)   & $M_1$         & $45.2^{+1.6}_{-1.5}$      & $0.32^{+0.09}_{-0.07}$    & $2.74^{+0.80}_{-0.53}$    & $1.45^{+0.40}_{-0.30}$    & $2.85^{+0.77}_{-0.60}$        \\
Mass, Secondary ($10^{18}$ kg) & $M_2$         & \nodata                   & $0.13^{+0.07}_{-0.09}$    & $1.31^{+0.53}_{-0.80}$    & $0.61^{+0.30}_{-0.40}$    & $1.19^{+0.60}_{-0.77}$        \\
Semi-major axis (km)           & $a$           & $1898^{+22}_{-21}$        & $8164^{+42}_{-42}$        & $9989^{+29}_{-29}$        & $7535^{+113}_{-115}$      & $9299^{+43}_{-39}$            \\
Eccentricity                   & $e$           & $0.036^{+0.017}_{-0.015}$ & $0.532^{+0.004}_{-0.004}$ & $0.352^{+0.002}_{-0.002}$ & $0.687^{+0.010}_{-0.009}$ & $0.552^{+0.005}_{-0.004}$     \\
Inclination ($^{\circ}$)       & $i$           & $65.8^{+1.9}_{-1.9}$      & $72.3^{+0.40}_{-0.37}$    & $25.11^{+0.23}_{-0.21}$   & $10.12^{+0.31}_{-0.36}$   & $20.24^{+0.20}_{-0.21}$       \\
Apsidal argument ($^{\circ}$)  & $\omega$      & $101^{+17}_{-24}$         & $201.99^{+0.32}_{-0.33}$  & $191.63^{+0.45}_{-0.43}$  & $170.4^{+0.8}_{-1.0}$     & $286.9^{+0.7}_{-0.7}$         \\
Nodal longitude ($^{\circ}$)   & $\Omega$      & $122.9^{+1.7}_{-1.6}$     & $17.09^{+0.33}_{-0.30}$   & $274.17^{+0.31}_{-0.32}$  & $13.3^{+1.0}_{-0.8}$      & $321.64^{+0.47}_{-0.47}$      \\
Mean anomaly ($^{\circ}$)      & $\mathcal{M}$ & $147^{+23}_{-17}$         & $34.11^{+0.39}_{-0.33}$   & $124.34^{+0.37}_{-0.37}$  & $312.67^{+0.46}_{-0.42}$  & $265.75^{+0.55}_{-0.57}$      \\
$J_2$ harmonic (km$^2$) & $J_2R^2$      & \nodata                   & $4258^{+5552}_{-3028}$    & $15755^{+8647}_{-6584}$   & $910^{+868}_{-614}$       & $11735^{+6229}_{-6228}$       \\
Axis obliquity ($^{\circ}$)    & $i_{sp}$      & \nodata                   & $72^{+35}_{-29}$          & $21^{+4}_{-4}$            & $44^{+26}_{-22}$          & $33^{+16}_{-10}$              \\
Axis precession ($^{\circ}$)   & $\Omega_{sp}$ & \nodata                   & $85^{+243}_{-61}$         & $243^{+17}_{-37}$         & $241^{+61}_{-98}$         & $316^{+7}_{-20}$              \\
Reference epoch (Julian Date) & & 2452400 & 2452600 & 2454300 & 2455600 & 2454300 \\
Primary diameter (km) & $D_1$ & $411\pm7$ & $82\pm18$ & $107\pm37$ & $138\pm24$ & $170\pm43$ \\
Secondary diameter (km) & $D_2$ & $213\pm30$ & $63\pm14$ & $96\pm34$ & $105\pm20$ & $141\pm36$ \\
\enddata
\tablecomments{All fitted angles are relative to the J2000 ecliptic plane. Reference epochs refer to the system-centric Julian date. Posteriors show the median, 16th, and 84th percentiles. For Huya, orbit solution uses a Keplerian orbit fit. For Logos-Zoe and \kh, no statistically significant non-Keplerian motion is detected, although the addition of these parameter allows for a more conservative propagation of uncertainties. Primary and secondary diameters are not used in any significant way in the orbit fitting process, but are used for predicting mutual events. Data used for orbit fitting is available on the \textit{Orbit Status of Known Binary TNOs} webpage (\url{www2.lowell.edu/users/grundy/tnbs/status.html}) as well as Table \ref{tab:observations}. Orbit solution sources are: Huya, \citet{rommel2025huya}; Altjira, \citet{nelsen2025beyond}; Logos-Zoe, \kh, \sssb{2001}{XR}{254}, this work. Primary and secondary diameter measurements are found from: Huya, \citet{fornasier2013tnos,santos2022physical,rommel2025huya}; Logos-Zoe, \citet{grundy2011five}; Altjira, \citet{vilenius2014tnos,nelsen2025beyond}; \kh, \citet{vilenius2012tnos}; \sssb{2001}{XR}{254}, \citet{vilenius2014tnos}. }
\label{tab:orbits}
\end{deluxetable*}

\section{Tables of Mutual Events}
\label{app:tables}

Below, in Tables \ref{tab:huya}--\ref{tab:xr} we show all mutual event predictions through the 2030s for events with event probability $\geq1\%$. For Huya, since there are nearly 2000 events, we display 20 selected mutual events early in the event season (Table \ref{tab:huya}). The complete table of mutual events of Huya are provided as a machine readable table.

\begin{longrotatetable}
\begin{deluxetable*}{lllllll}
\tabletypesize{\footnotesize}
\tablewidth{\textwidth}
\tablecaption{Selected Mutual Events of Huya}
\tablehead{
Event Mid-time & First Contact & Last Contact & Event Duration & Min. Separation & Median Depth & Prob. \\
 & & & (hours) & (mas) & (mag) & 
}
\startdata
2035-02-12 01:01 $\pm$ 0.69 hr & 2035-02-11 23:27 $\pm$ 0.83 hr & 2035-02-12 02:31 $\pm$ 0.68 hr & 3.23 $\pm$ 0.66 & 9.10 $\pm$ 2.12 & 0.10 $\pm$ 0.06 & 0.98 \\
2035-02-15 12:06 $\pm$ 0.69 hr & 2035-02-15 10:32 $\pm$ 0.82 hr & 2035-02-15 13:38 $\pm$ 0.68 hr & 3.27 $\pm$ 0.64 & 8.97 $\pm$ 2.12 & 0.10 $\pm$ 0.06 & 0.98 \\
2035-02-17 07:18 $\pm$ 0.64 hr & 2035-02-17 05:42 $\pm$ 0.71 hr & 2035-02-17 08:54 $\pm$ 0.78 hr & 3.30 $\pm$ 0.74 & 9.21 $\pm$ 2.17 & 0.10 $\pm$ 0.07 & 0.98 \\
2035-02-18 23:11 $\pm$ 0.69 hr & 2035-02-18 21:37 $\pm$ 0.82 hr & 2035-02-19 00:45 $\pm$ 0.68 hr & 3.30 $\pm$ 0.64 & 8.84 $\pm$ 2.12 & 0.11 $\pm$ 0.06 & 0.98 \\
2035-02-20 18:23 $\pm$ 0.64 hr & 2035-02-20 16:47 $\pm$ 0.70 hr & 2035-02-20 20:01 $\pm$ 0.77 hr & 3.37 $\pm$ 0.71 & 9.07 $\pm$ 2.17 & 0.10 $\pm$ 0.07 & 0.98 \\
2035-02-22 10:18 $\pm$ 0.69 hr & 2035-02-22 08:42 $\pm$ 0.82 hr & 2035-02-22 11:52 $\pm$ 0.68 hr & 3.33 $\pm$ 0.63 & 8.72 $\pm$ 2.12 & 0.11 $\pm$ 0.06 & 0.99 \\
2035-02-24 05:28 $\pm$ 0.64 hr & 2035-02-24 03:51 $\pm$ 0.70 hr & 2035-02-24 07:08 $\pm$ 0.77 hr & 3.38 $\pm$ 0.69 & 8.94 $\pm$ 2.18 & 0.10 $\pm$ 0.07 & 0.98 \\
2035-02-25 21:23 $\pm$ 0.69 hr & 2035-02-25 19:45 $\pm$ 0.83 hr & 2035-02-25 22:59 $\pm$ 0.67 hr & 3.37 $\pm$ 0.63 & 8.59 $\pm$ 2.13 & 0.12 $\pm$ 0.06 & 0.99 \\
2035-02-27 16:33 $\pm$ 0.64 hr & 2035-02-27 14:55 $\pm$ 0.70 hr & 2035-02-27 18:15 $\pm$ 0.77 hr & 3.43 $\pm$ 0.69 & 8.81 $\pm$ 2.18 & 0.11 $\pm$ 0.07 & 0.98 \\
2035-03-01 08:28 $\pm$ 0.69 hr & 2035-03-01 06:50 $\pm$ 0.83 hr & 2035-03-01 10:05 $\pm$ 0.67 hr & 3.40 $\pm$ 0.62 & 8.48 $\pm$ 2.13 & 0.12 $\pm$ 0.06 & 0.99 \\
2035-03-03 03:38 $\pm$ 0.64 hr & 2035-03-03 02:00 $\pm$ 0.69 hr & 2035-03-03 05:20 $\pm$ 0.76 hr & 3.43 $\pm$ 0.68 & 8.69 $\pm$ 2.18 & 0.11 $\pm$ 0.07 & 0.99 \\
2035-03-04 19:33 $\pm$ 0.69 hr & 2035-03-04 17:54 $\pm$ 0.82 hr & 2035-03-04 21:12 $\pm$ 0.67 hr & 3.43 $\pm$ 0.60 & 8.37 $\pm$ 2.13 & 0.12 $\pm$ 0.06 & 0.99 \\
2035-03-06 14:43 $\pm$ 0.64 hr & 2035-03-06 13:05 $\pm$ 0.69 hr & 2035-03-06 16:27 $\pm$ 0.76 hr & 3.47 $\pm$ 0.67 & 8.58 $\pm$ 2.19 & 0.12 $\pm$ 0.07 & 0.99 \\
2035-03-08 06:39 $\pm$ 0.69 hr & 2035-03-08 04:59 $\pm$ 0.81 hr & 2035-03-08 08:17 $\pm$ 0.67 hr & 3.43 $\pm$ 0.59 & 8.26 $\pm$ 2.14 & 0.13 $\pm$ 0.06 & 0.99 \\
2035-03-10 01:48 $\pm$ 0.64 hr & 2035-03-10 00:08 $\pm$ 0.69 hr & 2035-03-10 03:32 $\pm$ 0.76 hr & 3.50 $\pm$ 0.66 & 8.47 $\pm$ 2.19 & 0.12 $\pm$ 0.07 & 0.99 \\
2035-03-11 17:42 $\pm$ 0.69 hr & 2035-03-11 16:02 $\pm$ 0.81 hr & 2035-03-11 19:24 $\pm$ 0.67 hr & 3.47 $\pm$ 0.57 & 8.17 $\pm$ 2.14 & 0.13 $\pm$ 0.06 & 0.99 \\
2035-03-13 12:53 $\pm$ 0.64 hr & 2035-03-13 11:13 $\pm$ 0.68 hr & 2035-03-13 14:39 $\pm$ 0.76 hr & 3.53 $\pm$ 0.65 & 8.37 $\pm$ 2.20 & 0.12 $\pm$ 0.07 & 0.99 \\
2035-03-15 04:46 $\pm$ 0.69 hr & 2035-03-15 03:06 $\pm$ 0.82 hr & 2035-03-15 06:30 $\pm$ 0.67 hr & 3.47 $\pm$ 0.58 & 8.07 $\pm$ 2.14 & 0.14 $\pm$ 0.06 & 0.99 \\
2035-03-16 23:56 $\pm$ 0.64 hr & 2035-03-16 22:16 $\pm$ 0.68 hr & 2035-03-17 01:44 $\pm$ 0.75 hr & 3.57 $\pm$ 0.63 & 8.28 $\pm$ 2.20 & 0.13 $\pm$ 0.07 & 0.99 \\
2035-03-18 15:52 $\pm$ 0.69 hr & 2035-03-18 14:11 $\pm$ 0.82 hr & 2035-03-18 17:34 $\pm$ 0.67 hr & 3.50 $\pm$ 0.58 & 7.99 $\pm$ 2.15 & 0.14 $\pm$ 0.06 & 1.00 \\
\enddata
\tablecomments{All times are in UTC. First and last contact correspond to the start and end of dimming and are only calculated from statistical samples that produce mutual events. As seen in Figure \ref{fig:lc_bad}, it is possible for periods of no/minimal dimming between first and last contact. Event duration refers to the time between first and last contact. Median depth is calculated based on our highly simplified photometric model and is based on the whole set of samples (not just those that produce mutual events). Event probability is defined by the fraction of samples that produce mutual events.}
\label{tab:huya}
\end{deluxetable*}
\end{longrotatetable}

\begin{longrotatetable}
\begin{deluxetable*}{lllllll}
\tabletypesize{\footnotesize}
\tablewidth{\textwidth}
\tablecaption{Mutual Events of Logos-Zoe}
\tablehead{
Event Mid-time & First Contact & Last Contact & Event Duration & Min. Separation & Median Depth & Prob. \\
 & & & (hours) & (mas) & (mag) & 
}
\startdata
2027-05-14 19:16 $\pm$ 46.85 hr & 2027-05-14 15:27 $\pm$ 39.10 hr & 2027-05-14 22:43 $\pm$ 39.29 hr & 8.17 $\pm$ 3.47   & 3.80 $\pm$ 6.76 & 0.02 $\pm$ 0.08 & 0.65 \\
2027-08-27 21:02 $\pm$ 11.21 hr & 2027-08-27 08:02 $\pm$ 12.82 hr & 2027-08-28 08:04 $\pm$ 12.44 hr & 25.30 $\pm$ 10.13 & 8.64 $\pm$ 1.37 & 0.00 $\pm$ 0.08 & 0.17 \\
2028-03-19 09:07 $\pm$ 47.89 hr & 2028-03-19 03:21 $\pm$ 39.46 hr & 2028-03-19 15:47 $\pm$ 39.34 hr & 12.23 $\pm$ 2.57  & 1.02 $\pm$ 7.78 & 0.47 $\pm$ 0.16 & 0.91 \\
2028-07-02 20:53 $\pm$ 11.65 hr & 2028-07-02 06:48 $\pm$ 13.73 hr & 2028-07-03 10:19 $\pm$ 12.00 hr & 27.82 $\pm$ 9.54  & 5.17 $\pm$ 1.43 & 0.00 $\pm$ 0.15 & 0.49 \\
2029-01-22 22:44 $\pm$ 48.97 hr & 2029-01-22 19:54 $\pm$ 39.30 hr & 2029-01-23 01:43 $\pm$ 39.53 hr & 7.70 $\pm$ 3.22   & 4.85 $\pm$ 7.48 & 0.00 $\pm$ 0.06 & 0.49 \\
2029-05-08 07:18 $\pm$ 12.03 hr & 2029-05-07 21:36 $\pm$ 9.68 hr  & 2029-05-08 20:10 $\pm$ 9.90 hr  & 18.47 $\pm$ 11.11 & 4.91 $\pm$ 1.50 & 0.00 $\pm$ 0.03 & 0.04 \\
\enddata
\tablecomments{See Table \ref{tab:huya} for definitions and details of each column.}
\label{tab:logos}
\end{deluxetable*}
\end{longrotatetable}

\begin{longrotatetable}
\begin{deluxetable*}{lllllll}
\tabletypesize{\footnotesize}
\tablewidth{\textwidth}
\tablecaption{Mutual Events of Altjira}
\tablehead{
Event Mid-time & First Contact & Last Contact & Event Duration & Min. Separation & Median Depth & Prob. \\
 & & & (hours) & (mas) & (mag) & 
}
\startdata
2025-07-08 06:45 $\pm$ 18.71 hr & 2025-07-08 03:21 $\pm$ 19.47 hr & 2025-07-08 09:59 $\pm$ 17.61 hr & 8.13 $\pm$ 5.74   & 4.94 $\pm$ 0.98  & 0.00 $\pm$ 0.01 & 0.07 \\
2025-10-02 10:40 $\pm$ 0.71 hr  & 2025-10-02 04:17 $\pm$ 2.95 hr  & 2025-10-02 14:24 $\pm$ 1.42 hr  & 10.18 $\pm$ 4.07  & 0.89 $\pm$ 0.51  & 0.40 $\pm$ 0.18 & 0.96 \\
2025-11-25 04:14 $\pm$ 18.70 hr & 2025-11-24 23:42 $\pm$ 18.37 hr & 2025-11-25 09:10 $\pm$ 18.00 hr & 11.03 $\pm$ 5.58  & 3.04 $\pm$ 1.06  & 0.00 $\pm$ 0.11 & 0.51 \\
2026-02-18 23:20 $\pm$ 0.73 hr  & 2026-02-18 19:58 $\pm$ 2.76 hr  & 2026-02-19 02:40 $\pm$ 1.49 hr  & 6.67 $\pm$ 3.90   & 3.65 $\pm$ 0.55  & 0.12 $\pm$ 0.12 & 0.80 \\
2026-04-13 19:34 $\pm$ 19.30 hr & 2026-04-13 13:34 $\pm$ 18.81 hr & 2026-04-14 01:24 $\pm$ 18.40 hr & 11.05 $\pm$ 5.29  & 6.50 $\pm$ 1.04  & 0.00 $\pm$ 0.08 & 0.39 \\
2026-07-08 19:00 $\pm$ 0.75 hr  & 2026-07-08 13:35 $\pm$ 1.80 hr  & 2026-07-08 22:56 $\pm$ 1.44 hr  & 9.38 $\pm$ 2.86   & 0.87 $\pm$ 0.52  & 0.41 $\pm$ 0.18 & 0.96 \\
2026-09-01 07:18 $\pm$ 20.16 hr & 2026-08-31 17:58 $\pm$ 22.34 hr & 2026-09-01 17:24 $\pm$ 21.29 hr & 20.93 $\pm$ 10.63 & 1.29 $\pm$ 0.87  & 0.39 $\pm$ 0.18 & 0.97 \\
2026-11-25 10:58 $\pm$ 0.77 hr  & 2026-11-25 06:16 $\pm$ 1.70 hr  & 2026-11-25 15:17 $\pm$ 1.52 hr  & 8.90 $\pm$ 2.80   & 0.47 $\pm$ 0.37  & 0.62 $\pm$ 0.11 & 1.00 \\
2027-01-18 14:27 $\pm$ 19.98 hr & 2027-01-18 02:31 $\pm$ 21.30 hr & 2027-01-18 22:15 $\pm$ 20.26 hr & 22.68 $\pm$ 8.73  & 2.25 $\pm$ 1.11  & 0.42 $\pm$ 0.22 & 0.94 \\
2027-04-14 03:30 $\pm$ 0.80 hr  & 2027-04-13 20:26 $\pm$ 2.50 hr  & 2027-04-14 07:24 $\pm$ 1.51 hr  & 10.87 $\pm$ 3.57  & 1.67 $\pm$ 0.58  & 0.56 $\pm$ 0.14 & 0.99 \\
2027-06-07 12:38 $\pm$ 21.10 hr & 2027-06-07 06:10 $\pm$ 21.04 hr & 2027-06-07 19:52 $\pm$ 20.96 hr & 14.63 $\pm$ 5.20  & 0.78 $\pm$ 0.73  & 0.61 $\pm$ 0.22 & 0.98 \\
2027-08-31 20:57 $\pm$ 0.82 hr  & 2027-08-31 17:19 $\pm$ 1.63 hr  & 2027-09-01 04:07 $\pm$ 3.04 hr  & 10.85 $\pm$ 4.26  & 2.41 $\pm$ 0.61  & 0.46 $\pm$ 0.18 & 0.98 \\
2027-10-25 10:09 $\pm$ 21.38 hr & 2027-10-25 04:43 $\pm$ 21.59 hr & 2027-10-25 17:05 $\pm$ 22.66 hr & 13.37 $\pm$ 7.46  & 5.17 $\pm$ 1.19  & 0.17 $\pm$ 0.20 & 0.76 \\
2028-01-18 11:22 $\pm$ 0.84 hr  & 2028-01-18 07:32 $\pm$ 1.64 hr  & 2028-01-18 17:42 $\pm$ 2.03 hr  & 10.10 $\pm$ 3.24  & 0.72 $\pm$ 0.50  & 0.49 $\pm$ 0.18 & 0.98 \\
2028-03-12 15:57 $\pm$ 21.59 hr & 2028-03-12 07:45 $\pm$ 22.29 hr & 2028-03-13 04:23 $\pm$ 24.00 hr & 15.68 $\pm$ 10.13 & 0.81 $\pm$ 0.74  & 0.43 $\pm$ 0.19 & 0.97 \\
2028-06-06 07:23 $\pm$ 0.87 hr  & 2028-06-06 03:55 $\pm$ 1.66 hr  & 2028-06-06 10:57 $\pm$ 1.64 hr  & 6.97 $\pm$ 2.80   & 1.50 $\pm$ 0.61  & 0.23 $\pm$ 0.18 & 0.90 \\
2028-07-31 10:19 $\pm$ 22.75 hr & 2028-07-31 02:39 $\pm$ 25.31 hr & 2028-07-31 19:35 $\pm$ 26.71 hr & 11.67 $\pm$ 6.52  & 6.55 $\pm$ 1.19  & 0.00 $\pm$ 0.05 & 0.15 \\
2028-10-24 01:04 $\pm$ 0.90 hr  & 2028-10-23 21:54 $\pm$ 1.68 hr  & 2028-10-24 04:06 $\pm$ 2.04 hr  & 6.10 $\pm$ 3.25   & 4.54 $\pm$ 0.66  & 0.02 $\pm$ 0.09 & 0.57 \\
2028-12-18 09:27 $\pm$ 22.54 hr & 2028-12-18 02:47 $\pm$ 27.65 hr & 2028-12-18 16:05 $\pm$ 28.19 hr & 11.23 $\pm$ 4.61  & 6.57 $\pm$ 1.28  & 0.00 $\pm$ 0.02 & 0.05 \\
2029-03-12 13:29 $\pm$ 0.92 hr  & 2029-03-12 10:07 $\pm$ 1.70 hr  & 2029-03-12 17:19 $\pm$ 2.98 hr  & 6.83 $\pm$ 4.20   & 1.70 $\pm$ 0.65  & 0.20 $\pm$ 0.17 & 0.87 \\
2029-05-06 16:24 $\pm$ 23.42 hr & 2029-05-06 10:42 $\pm$ 25.46 hr & 2029-05-06 22:10 $\pm$ 25.73 hr & 11.77 $\pm$ 5.82  & 4.47 $\pm$ 1.24  & 0.00 $\pm$ 0.06 & 0.18 \\
2029-07-30 11:54 $\pm$ 0.96 hr  & 2029-07-30 09:04 $\pm$ 1.59 hr  & 2029-07-30 14:28 $\pm$ 2.15 hr  & 5.77 $\pm$ 3.24   & 5.08 $\pm$ 0.68  & 0.00 $\pm$ 0.03 & 0.17 \\
2029-12-17 04:41 $\pm$ 0.98 hr  & 2029-12-17 02:24 $\pm$ 1.55 hr  & 2029-12-17 07:21 $\pm$ 1.54 hr  & 4.73 $\pm$ 2.50   & 5.34 $\pm$ 0.72  & 0.00 $\pm$ 0.02 & 0.09 \\
2030-05-05 19:29 $\pm$ 1.01 hr  & 2030-05-05 16:45 $\pm$ 1.67 hr  & 2030-05-05 22:09 $\pm$ 1.67 hr  & 5.47 $\pm$ 2.73   & 4.04 $\pm$ 0.70  & 0.00 $\pm$ 0.03 & 0.20 \\
\enddata
\tablecomments{See Table \ref{tab:huya} for definitions and details of each column.}
\label{tab:altjira}
\end{deluxetable*}
\end{longrotatetable}

\begin{longrotatetable}
\begin{deluxetable*}{lllllll}
\tabletypesize{\footnotesize}
\tablewidth{\textwidth}
\tablecaption{Mutual Events of \kh}
\tablehead{
Event Mid-time & First Contact & Last Contact & Event Duration & Min. Separation & Median Depth & Prob. \\
 & & & (hours) & (mas) & (mag) & 
}
\startdata
2025-03-10 14:07 $\pm$ 5.78 hr  & 2025-03-10 10:47 $\pm$ 5.68 hr  & 2025-03-10 17:53 $\pm$ 5.93 hr  & 7.00 $\pm$ 1.24   & 0.27 $\pm$ 0.26  & 0.49 $\pm$ 0.05 & 1.00 \\
2025-05-05 19:54 $\pm$ 12.30 hr & 2025-05-04 15:16 $\pm$ 13.34 hr & 2025-05-07 10:48 $\pm$ 13.74 hr & 64.90 $\pm$ 14.15 & 1.34 $\pm$ 1.10  & 0.44 $\pm$ 0.14 & 0.98 \\
2025-07-16 19:10 $\pm$ 5.95 hr  & 2025-07-16 15:09 $\pm$ 5.86 hr  & 2025-07-16 23:22 $\pm$ 6.15 hr  & 8.13 $\pm$ 1.23   & 0.27 $\pm$ 0.21  & 0.49 $\pm$ 0.05 & 1.00 \\
2025-09-12 10:49 $\pm$ 12.97 hr & 2025-09-11 14:19 $\pm$ 13.05 hr & 2025-09-13 07:53 $\pm$ 14.22 hr & 41.03 $\pm$ 9.09  & 1.19 $\pm$ 1.24  & 0.49 $\pm$ 0.27 & 0.94 \\
2025-11-22 01:47 $\pm$ 6.27 hr  & 2025-11-21 22:20 $\pm$ 6.12 hr  & 2025-11-22 06:20 $\pm$ 6.42 hr  & 7.97 $\pm$ 1.24   & 0.62 $\pm$ 0.32  & 0.48 $\pm$ 0.05 & 1.00 \\
2026-01-19 13:11 $\pm$ 12.13 hr & 2026-01-18 21:35 $\pm$ 9.35 hr  & 2026-01-21 04:41 $\pm$ 23.51 hr & 42.05 $\pm$ 21.75 & 4.34 $\pm$ 1.82  & 0.18 $\pm$ 0.18 & 0.81 \\
2026-03-30 06:55 $\pm$ 6.42 hr  & 2026-03-30 03:25 $\pm$ 6.32 hr  & 2026-03-30 10:22 $\pm$ 6.59 hr  & 6.98 $\pm$ 1.25   & 0.57 $\pm$ 0.34  & 0.48 $\pm$ 0.06 & 1.00 \\
2026-05-26 16:13 $\pm$ 13.10 hr & 2026-05-25 23:48 $\pm$ 12.95 hr & 2026-05-28 21:16 $\pm$ 22.16 hr & 54.13 $\pm$ 20.82 & 1.74 $\pm$ 1.60  & 0.31 $\pm$ 0.19 & 0.88 \\
2026-08-05 11:45 $\pm$ 6.62 hr  & 2026-08-05 08:16 $\pm$ 6.53 hr  & 2026-08-05 16:03 $\pm$ 6.81 hr  & 7.77 $\pm$ 1.24   & 0.35 $\pm$ 0.30  & 0.48 $\pm$ 0.05 & 1.00 \\
2026-10-04 14:43 $\pm$ 13.44 hr & 2026-10-03 22:04 $\pm$ 13.58 hr & 2026-10-05 08:29 $\pm$ 13.20 hr & 35.10 $\pm$ 12.27 & 3.80 $\pm$ 1.82  & 0.01 $\pm$ 0.18 & 0.54 \\
2026-12-11 18:31 $\pm$ 6.94 hr  & 2026-12-11 15:03 $\pm$ 6.78 hr  & 2026-12-11 23:05 $\pm$ 7.09 hr  & 8.03 $\pm$ 1.28   & 1.14 $\pm$ 0.34  & 0.45 $\pm$ 0.07 & 1.00 \\
2027-02-10 07:32 $\pm$ 12.66 hr & 2027-02-09 19:02 $\pm$ 11.64 hr & 2027-02-11 01:52 $\pm$ 18.10 hr & 25.95 $\pm$ 18.33 & 6.56 $\pm$ 1.96  & 0.00 $\pm$ 0.09 & 0.38 \\
2027-04-18 23:12 $\pm$ 7.07 hr  & 2027-04-18 19:40 $\pm$ 6.98 hr  & 2027-04-19 03:00 $\pm$ 7.26 hr  & 7.33 $\pm$ 1.28   & 0.92 $\pm$ 0.36  & 0.45 $\pm$ 0.07 & 1.00 \\
2027-06-17 05:36 $\pm$ 13.90 hr & 2027-06-16 16:55 $\pm$ 14.02 hr & 2027-06-18 03:16 $\pm$ 24.25 hr & 27.42 $\pm$ 20.40 & 3.75 $\pm$ 1.96  & 0.02 $\pm$ 0.14 & 0.59 \\
2027-08-25 04:49 $\pm$ 7.31 hr  & 2027-08-25 01:21 $\pm$ 7.21 hr  & 2027-08-25 08:36 $\pm$ 7.48 hr  & 7.23 $\pm$ 1.26   & 0.85 $\pm$ 0.35  & 0.45 $\pm$ 0.07 & 1.00 \\
2027-10-26 02:14 $\pm$ 13.84 hr & 2027-10-25 16:19 $\pm$ 12.43 hr & 2027-10-27 01:09 $\pm$ 20.43 hr & 31.55 $\pm$ 18.02 & 6.86 $\pm$ 1.95  & 0.00 $\pm$ 0.04 & 0.13 \\
2027-12-31 11:06 $\pm$ 7.61 hr  & 2027-12-31 07:37 $\pm$ 7.46 hr  & 2027-12-31 15:33 $\pm$ 7.76 hr  & 7.83 $\pm$ 1.35   & 1.64 $\pm$ 0.36  & 0.38 $\pm$ 0.08 & 1.00 \\
2028-03-03 00:52 $\pm$ 13.25 hr & 2028-03-02 13:53 $\pm$ 11.42 hr & 2028-03-03 09:56 $\pm$ 12.81 hr & 21.22 $\pm$ 9.43  & 8.60 $\pm$ 2.08  & 0.00 $\pm$ 0.03 & 0.06 \\
2028-05-07 15:30 $\pm$ 7.74 hr  & 2028-05-07 11:51 $\pm$ 7.64 hr  & 2028-05-07 19:36 $\pm$ 7.93 hr  & 7.53 $\pm$ 1.33   & 1.26 $\pm$ 0.38  & 0.39 $\pm$ 0.08 & 1.00 \\
2028-07-08 00:17 $\pm$ 14.66 hr & 2028-07-07 11:37 $\pm$ 14.14 hr & 2028-07-08 14:13 $\pm$ 19.29 hr & 24.65 $\pm$ 12.12 & 6.03 $\pm$ 2.07  & 0.00 $\pm$ 0.05 & 0.17 \\
2028-09-12 21:45 $\pm$ 8.01 hr  & 2028-09-12 18:25 $\pm$ 7.91 hr  & 2028-09-13 01:09 $\pm$ 8.15 hr  & 6.50 $\pm$ 1.33   & 1.37 $\pm$ 0.37  & 0.35 $\pm$ 0.08 & 1.00 \\
2029-01-19 03:44 $\pm$ 8.28 hr  & 2029-01-19 00:10 $\pm$ 8.15 hr  & 2029-01-19 07:51 $\pm$ 8.42 hr  & 7.40 $\pm$ 1.48   & 2.10 $\pm$ 0.38  & 0.30 $\pm$ 0.09 & 1.00 \\
2029-05-27 07:57 $\pm$ 8.41 hr  & 2029-05-27 04:12 $\pm$ 8.32 hr  & 2029-05-27 12:06 $\pm$ 8.61 hr  & 7.57 $\pm$ 1.42   & 1.61 $\pm$ 0.40  & 0.32 $\pm$ 0.09 & 1.00 \\
2029-10-02 14:48 $\pm$ 8.71 hr  & 2029-10-02 11:38 $\pm$ 8.62 hr  & 2029-10-02 17:50 $\pm$ 8.85 hr  & 5.90 $\pm$ 1.45   & 1.91 $\pm$ 0.38  & 0.24 $\pm$ 0.09 & 1.00 \\
2030-02-07 20:13 $\pm$ 8.95 hr  & 2030-02-07 16:49 $\pm$ 8.82 hr  & 2030-02-07 23:43 $\pm$ 9.10 hr  & 6.70 $\pm$ 1.63   & 2.54 $\pm$ 0.39  & 0.22 $\pm$ 0.09 & 0.99 \\
2030-06-16 00:25 $\pm$ 9.09 hr  & 2030-06-15 20:49 $\pm$ 9.01 hr  & 2030-06-16 04:13 $\pm$ 9.30 hr  & 7.30 $\pm$ 1.60   & 1.99 $\pm$ 0.41  & 0.25 $\pm$ 0.09 & 1.00 \\
2030-10-22 07:40 $\pm$ 9.41 hr  & 2030-10-22 04:28 $\pm$ 9.16 hr  & 2030-10-22 10:27 $\pm$ 9.44 hr  & 5.83 $\pm$ 1.56   & 2.46 $\pm$ 0.39  & 0.17 $\pm$ 0.08 & 0.97 \\
2031-02-27 12:38 $\pm$ 9.63 hr  & 2031-02-27 09:38 $\pm$ 9.32 hr  & 2031-02-27 15:38 $\pm$ 9.65 hr  & 5.90 $\pm$ 1.70   & 2.95 $\pm$ 0.41  & 0.15 $\pm$ 0.08 & 0.97 \\
2031-07-05 17:03 $\pm$ 9.79 hr  & 2031-07-05 13:41 $\pm$ 9.71 hr  & 2031-07-05 20:33 $\pm$ 9.98 hr  & 6.80 $\pm$ 1.81   & 2.39 $\pm$ 0.42  & 0.18 $\pm$ 0.09 & 0.98 \\
2031-11-11 00:32 $\pm$ 10.11 hr & 2031-11-10 21:44 $\pm$ 9.86 hr  & 2031-11-11 03:26 $\pm$ 10.10 hr & 5.67 $\pm$ 1.70   & 3.02 $\pm$ 0.40  & 0.10 $\pm$ 0.08 & 0.92 \\
2032-03-18 05:28 $\pm$ 10.30 hr & 2032-03-18 03:00 $\pm$ 10.10 hr & 2032-03-18 07:53 $\pm$ 10.29 hr & 4.90 $\pm$ 1.69   & 3.33 $\pm$ 0.43  & 0.09 $\pm$ 0.07 & 0.92 \\
2032-07-24 09:48 $\pm$ 10.49 hr & 2032-07-24 07:04 $\pm$ 10.25 hr & 2032-07-24 12:50 $\pm$ 10.55 hr & 6.07 $\pm$ 1.86   & 2.82 $\pm$ 0.43  & 0.11 $\pm$ 0.08 & 0.93 \\
2032-11-29 17:01 $\pm$ 10.81 hr & 2032-11-29 14:45 $\pm$ 10.60 hr & 2032-11-29 19:41 $\pm$ 10.89 hr & 5.30 $\pm$ 1.94   & 3.58 $\pm$ 0.42  & 0.04 $\pm$ 0.06 & 0.80 \\
2033-04-06 22:02 $\pm$ 10.97 hr & 2033-04-06 20:06 $\pm$ 10.79 hr & 2033-04-07 00:08 $\pm$ 11.00 hr & 4.33 $\pm$ 1.71   & 3.67 $\pm$ 0.45  & 0.04 $\pm$ 0.05 & 0.79 \\
2033-08-13 02:47 $\pm$ 11.19 hr & 2033-08-13 00:33 $\pm$ 11.01 hr & 2033-08-13 05:19 $\pm$ 11.21 hr & 5.10 $\pm$ 2.01   & 3.28 $\pm$ 0.44  & 0.05 $\pm$ 0.06 & 0.83 \\
2033-12-19 09:19 $\pm$ 11.51 hr & 2033-12-19 07:26 $\pm$ 11.29 hr & 2033-12-19 11:53 $\pm$ 11.46 hr & 4.78 $\pm$ 2.15   & 4.13 $\pm$ 0.44  & 0.01 $\pm$ 0.04 & 0.64 \\
2034-04-26 14:17 $\pm$ 11.65 hr & 2034-04-26 12:13 $\pm$ 11.29 hr & 2034-04-26 16:10 $\pm$ 11.49 hr & 4.37 $\pm$ 1.80   & 4.01 $\pm$ 0.46  & 0.01 $\pm$ 0.04 & 0.62 \\
2034-09-01 19:31 $\pm$ 11.90 hr & 2034-09-01 17:43 $\pm$ 11.62 hr & 2034-09-01 21:31 $\pm$ 11.77 hr & 4.03 $\pm$ 1.87   & 3.77 $\pm$ 0.45  & 0.01 $\pm$ 0.04 & 0.63 \\
2035-01-08 01:58 $\pm$ 12.20 hr & 2035-01-08 00:38 $\pm$ 11.91 hr & 2035-01-08 04:10 $\pm$ 12.20 hr & 4.35 $\pm$ 2.17   & 4.67 $\pm$ 0.45  & 0.00 $\pm$ 0.03 & 0.43 \\
2035-05-16 06:37 $\pm$ 12.33 hr & 2035-05-16 05:01 $\pm$ 12.06 hr & 2035-05-16 09:17 $\pm$ 12.31 hr & 4.53 $\pm$ 1.89   & 4.34 $\pm$ 0.48  & 0.00 $\pm$ 0.03 & 0.43 \\
2035-09-21 12:57 $\pm$ 12.61 hr & 2035-09-21 11:45 $\pm$ 12.01 hr & 2035-09-21 14:27 $\pm$ 12.21 hr & 3.52 $\pm$ 1.67   & 4.28 $\pm$ 0.46  & 0.00 $\pm$ 0.02 & 0.37 \\
2036-01-27 19:45 $\pm$ 12.89 hr & 2036-01-27 18:03 $\pm$ 12.87 hr & 2036-01-27 21:27 $\pm$ 13.05 hr & 3.40 $\pm$ 1.82   & 5.17 $\pm$ 0.47  & 0.00 $\pm$ 0.02 & 0.24 \\
2036-06-03 23:43 $\pm$ 13.02 hr & 2036-06-03 22:05 $\pm$ 12.94 hr & 2036-06-04 01:41 $\pm$ 13.22 hr & 4.50 $\pm$ 1.90   & 4.67 $\pm$ 0.49  & 0.00 $\pm$ 0.02 & 0.27 \\
2036-10-10 07:30 $\pm$ 13.33 hr & 2036-10-10 06:02 $\pm$ 13.38 hr & 2036-10-10 08:50 $\pm$ 13.43 hr & 2.95 $\pm$ 1.51   & 4.83 $\pm$ 0.47  & 0.00 $\pm$ 0.01 & 0.18 \\
2037-02-15 12:26 $\pm$ 13.57 hr & 2037-02-15 11:05 $\pm$ 13.11 hr & 2037-02-15 13:56 $\pm$ 13.00 hr & 2.88 $\pm$ 1.73   & 5.64 $\pm$ 0.49  & 0.00 $\pm$ 0.01 & 0.13 \\
2037-06-23 16:52 $\pm$ 13.72 hr & 2037-06-23 15:34 $\pm$ 13.88 hr & 2037-06-23 18:42 $\pm$ 13.93 hr & 3.07 $\pm$ 2.00   & 5.02 $\pm$ 0.51  & 0.00 $\pm$ 0.01 & 0.17 \\
2037-10-29 23:20 $\pm$ 14.04 hr & 2037-10-29 21:48 $\pm$ 14.00 hr & 2037-10-30 01:36 $\pm$ 13.68 hr & 2.57 $\pm$ 1.59   & 5.41 $\pm$ 0.49  & 0.00 $\pm$ 0.00 & 0.05 \\
2038-03-07 03:59 $\pm$ 14.26 hr & 2038-03-07 02:32 $\pm$ 14.47 hr & 2038-03-07 05:31 $\pm$ 14.11 hr & 2.35 $\pm$ 1.55   & 6.07 $\pm$ 0.52  & 0.00 $\pm$ 0.00 & 0.05 \\
2038-07-13 06:04 $\pm$ 14.43 hr & 2038-07-13 04:58 $\pm$ 13.35 hr & 2038-07-13 07:10 $\pm$ 13.34 hr & 2.73 $\pm$ 1.88   & 5.41 $\pm$ 0.52  & 0.00 $\pm$ 0.01 & 0.07 \\
\enddata
\tablecomments{See Table \ref{tab:huya} for definitions and details of each column.}
\label{tab:kh}
\end{deluxetable*}
\end{longrotatetable}

\begin{longrotatetable}
\begin{deluxetable*}{lllllll}
\tabletypesize{\footnotesize}
\tablewidth{\textwidth}
\tablecaption{Mutual Events of XR254}
\tablehead{
Event Mid-time & First Contact & Last Contact & Event Duration & Min. Separation & Median Depth & Prob. \\
 & & & (hours) & (mas) & (mag) & 
}
\startdata
2031-09-25 23:25 $\pm$ 5.21 hr  & 2031-09-25 19:47 $\pm$ 5.18 hr  & 2031-09-26 03:15 $\pm$ 4.59 hr  & 6.53 $\pm$ 3.52  & 8.99 $\pm$ 3.11  & 0.00 $\pm$ 0.02 & 0.11 \\
2032-01-29 13:18 $\pm$ 5.25 hr  & 2032-01-29 09:16 $\pm$ 4.95 hr  & 2032-01-29 17:10 $\pm$ 4.44 hr  & 7.10 $\pm$ 3.13  & 9.73 $\pm$ 3.27  & 0.00 $\pm$ 0.02 & 0.08 \\
2032-06-03 06:13 $\pm$ 5.29 hr  & 2032-06-03 02:18 $\pm$ 5.57 hr  & 2032-06-03 09:33 $\pm$ 4.72 hr  & 6.42 $\pm$ 3.86  & 10.30 $\pm$ 3.18 & 0.00 $\pm$ 0.02 & 0.10 \\
2032-09-13 15:33 $\pm$ 20.73 hr & 2032-09-13 13:09 $\pm$ 19.02 hr & 2032-09-13 23:17 $\pm$ 17.33 hr & 9.97 $\pm$ 5.77  & 13.20 $\pm$ 5.66 & 0.00 $\pm$ 0.02 & 0.05 \\
2032-10-07 01:39 $\pm$ 5.54 hr  & 2032-10-06 20:51 $\pm$ 5.87 hr  & 2032-10-07 06:05 $\pm$ 5.68 hr  & 8.88 $\pm$ 4.12  & 7.73 $\pm$ 3.28  & 0.00 $\pm$ 0.07 & 0.22 \\
2033-01-17 06:49 $\pm$ 20.72 hr & 2033-01-17 02:23 $\pm$ 18.51 hr & 2033-01-17 12:35 $\pm$ 17.18 hr & 10.30 $\pm$ 6.01 & 13.49 $\pm$ 5.93 & 0.00 $\pm$ 0.03 & 0.06 \\
2033-02-09 15:07 $\pm$ 5.54 hr  & 2033-02-09 11:13 $\pm$ 5.76 hr  & 2033-02-09 18:59 $\pm$ 5.56 hr  & 7.73 $\pm$ 3.36  & 8.76 $\pm$ 3.42  & 0.00 $\pm$ 0.05 & 0.18 \\
2033-06-15 07:54 $\pm$ 5.60 hr  & 2033-06-15 02:54 $\pm$ 6.24 hr  & 2033-06-15 11:50 $\pm$ 5.61 hr  & 9.10 $\pm$ 4.20  & 8.98 $\pm$ 3.32  & 0.00 $\pm$ 0.07 & 0.20 \\
2033-09-26 00:21 $\pm$ 20.86 hr & 2033-09-25 17:41 $\pm$ 21.52 hr & 2033-09-26 07:35 $\pm$ 21.23 hr & 16.37 $\pm$ 6.79 & 10.78 $\pm$ 5.75 & 0.00 $\pm$ 0.08 & 0.15 \\
2033-10-19 04:18 $\pm$ 5.89 hr  & 2033-10-18 21:55 $\pm$ 6.00 hr  & 2033-10-19 09:09 $\pm$ 5.73 hr  & 11.57 $\pm$ 4.04 & 6.47 $\pm$ 3.42  & 0.00 $\pm$ 0.15 & 0.32 \\
2034-01-29 13:49 $\pm$ 21.13 hr & 2034-01-29 09:11 $\pm$ 21.80 hr & 2034-01-29 20:31 $\pm$ 21.37 hr & 14.40 $\pm$ 5.74 & 11.73 $\pm$ 6.06 & 0.00 $\pm$ 0.07 & 0.14 \\
2034-02-21 17:08 $\pm$ 5.86 hr  & 2034-02-21 12:36 $\pm$ 5.75 hr  & 2034-02-21 21:34 $\pm$ 5.83 hr  & 8.83 $\pm$ 3.62  & 7.80 $\pm$ 3.56  & 0.00 $\pm$ 0.11 & 0.30 \\
2034-06-04 06:16 $\pm$ 22.46 hr & 2034-06-03 22:13 $\pm$ 22.57 hr & 2034-06-04 13:20 $\pm$ 21.63 hr & 16.80 $\pm$ 7.67 & 13.22 $\pm$ 6.09 & 0.00 $\pm$ 0.08 & 0.15 \\
2034-06-27 10:01 $\pm$ 5.94 hr  & 2034-06-27 04:08 $\pm$ 5.95 hr  & 2034-06-27 14:28 $\pm$ 5.85 hr  & 10.67 $\pm$ 3.89 & 7.64 $\pm$ 3.46  & 0.00 $\pm$ 0.14 & 0.30 \\
2034-10-08 08:04 $\pm$ 20.97 hr & 2034-10-07 21:41 $\pm$ 22.51 hr & 2034-10-08 16:22 $\pm$ 20.92 hr & 19.08 $\pm$ 7.27 & 8.48 $\pm$ 5.73  & 0.00 $\pm$ 0.16 & 0.29 \\
2034-10-31 06:25 $\pm$ 6.27 hr  & 2034-10-30 23:31 $\pm$ 6.65 hr  & 2034-10-31 11:27 $\pm$ 6.20 hr  & 12.60 $\pm$ 4.88 & 5.20 $\pm$ 3.48  & 0.00 $\pm$ 0.23 & 0.47 \\
2035-02-10 18:35 $\pm$ 21.54 hr & 2035-02-10 10:49 $\pm$ 23.22 hr & 2035-02-11 02:04 $\pm$ 22.41 hr & 14.50 $\pm$ 4.99 & 10.02 $\pm$ 6.14 & 0.00 $\pm$ 0.13 & 0.24 \\
2035-03-05 18:53 $\pm$ 6.18 hr  & 2035-03-05 13:57 $\pm$ 6.28 hr  & 2035-03-05 23:52 $\pm$ 6.26 hr  & 10.88 $\pm$ 4.06 & 6.78 $\pm$ 3.68  & 0.00 $\pm$ 0.20 & 0.39 \\
2035-06-16 15:24 $\pm$ 22.63 hr & 2035-06-16 03:07 $\pm$ 23.79 hr & 2035-06-16 21:01 $\pm$ 21.60 hr & 19.30 $\pm$ 8.01 & 10.83 $\pm$ 6.13 & 0.00 $\pm$ 0.15 & 0.28 \\
2035-07-09 11:41 $\pm$ 6.31 hr  & 2035-07-09 05:59 $\pm$ 6.48 hr  & 2035-07-09 16:44 $\pm$ 6.41 hr  & 11.67 $\pm$ 4.27 & 6.26 $\pm$ 3.57  & 0.00 $\pm$ 0.23 & 0.41 \\
2035-10-20 10:12 $\pm$ 21.07 hr & 2035-10-20 02:10 $\pm$ 22.25 hr & 2035-10-20 17:48 $\pm$ 20.20 hr & 21.50 $\pm$ 8.18 & 6.26 $\pm$ 5.34  & 0.00 $\pm$ 0.22 & 0.42 \\
2035-11-12 07:45 $\pm$ 6.66 hr  & 2035-11-12 00:50 $\pm$ 6.51 hr  & 2035-11-12 13:04 $\pm$ 6.87 hr  & 12.95 $\pm$ 4.48 & 3.94 $\pm$ 3.27  & 0.06 $\pm$ 0.27 & 0.60 \\
2036-02-22 21:59 $\pm$ 21.92 hr & 2036-02-22 14:16 $\pm$ 22.72 hr & 2036-02-23 08:02 $\pm$ 21.74 hr & 16.77 $\pm$ 5.57 & 8.29 $\pm$ 6.05  & 0.00 $\pm$ 0.19 & 0.35 \\
2036-03-16 20:49 $\pm$ 6.53 hr  & 2036-03-16 15:14 $\pm$ 6.59 hr  & 2036-03-17 01:44 $\pm$ 6.77 hr  & 11.22 $\pm$ 4.34 & 5.70 $\pm$ 3.73  & 0.01 $\pm$ 0.27 & 0.53 \\
2036-06-27 17:05 $\pm$ 22.75 hr & 2036-06-27 06:21 $\pm$ 23.98 hr & 2036-06-28 02:01 $\pm$ 21.47 hr & 20.50 $\pm$ 7.66 & 8.45 $\pm$ 6.00  & 0.00 $\pm$ 0.21 & 0.38 \\
2036-07-20 13:27 $\pm$ 6.71 hr  & 2036-07-20 07:59 $\pm$ 6.57 hr  & 2036-07-20 18:19 $\pm$ 6.86 hr  & 10.93 $\pm$ 3.97 & 4.82 $\pm$ 3.51  & 0.02 $\pm$ 0.28 & 0.55 \\
2036-10-31 13:20 $\pm$ 21.17 hr & 2036-10-31 05:01 $\pm$ 21.35 hr & 2036-10-31 22:18 $\pm$ 20.01 hr & 20.90 $\pm$ 7.92 & 4.64 $\pm$ 4.61  & 0.04 $\pm$ 0.22 & 0.55 \\
2036-11-23 08:31 $\pm$ 7.06 hr  & 2036-11-23 01:53 $\pm$ 6.74 hr  & 2036-11-23 14:39 $\pm$ 7.36 hr  & 13.53 $\pm$ 4.51 & 3.03 $\pm$ 2.88  & 0.19 $\pm$ 0.27 & 0.70 \\
2037-03-06 02:03 $\pm$ 22.27 hr & 2037-03-05 18:37 $\pm$ 22.02 hr & 2037-03-06 09:01 $\pm$ 20.83 hr & 17.97 $\pm$ 6.14 & 6.56 $\pm$ 5.69  & 0.00 $\pm$ 0.21 & 0.46 \\
2037-03-28 22:06 $\pm$ 6.90 hr  & 2037-03-28 15:47 $\pm$ 6.84 hr  & 2037-03-29 03:38 $\pm$ 7.36 hr  & 12.20 $\pm$ 4.12 & 4.60 $\pm$ 3.60  & 0.10 $\pm$ 0.29 & 0.63 \\
2037-07-09 17:56 $\pm$ 22.80 hr & 2037-07-09 10:32 $\pm$ 22.30 hr & 2037-07-10 01:30 $\pm$ 21.03 hr & 18.70 $\pm$ 7.04 & 6.07 $\pm$ 5.43  & 0.00 $\pm$ 0.21 & 0.50 \\
2037-08-01 15:07 $\pm$ 7.14 hr  & 2037-08-01 09:59 $\pm$ 7.01 hr  & 2037-08-01 20:19 $\pm$ 7.46 hr  & 10.83 $\pm$ 3.54 & 3.44 $\pm$ 3.12  & 0.11 $\pm$ 0.26 & 0.65 \\
2037-11-12 15:26 $\pm$ 21.26 hr & 2037-11-12 04:14 $\pm$ 21.25 hr & 2037-11-13 00:12 $\pm$ 20.90 hr & 20.70 $\pm$ 7.75 & 4.35 $\pm$ 3.93  & 0.09 $\pm$ 0.22 & 0.63 \\
2037-12-05 10:20 $\pm$ 7.49 hr  & 2037-12-05 03:40 $\pm$ 7.23 hr  & 2037-12-05 16:16 $\pm$ 7.47 hr  & 13.53 $\pm$ 4.44 & 2.99 $\pm$ 2.55  & 0.24 $\pm$ 0.26 & 0.78 \\
2038-03-18 03:47 $\pm$ 22.59 hr & 2038-03-17 20:01 $\pm$ 22.15 hr & 2038-03-18 12:53 $\pm$ 21.30 hr & 18.10 $\pm$ 6.31 & 5.10 $\pm$ 5.09  & 0.03 $\pm$ 0.22 & 0.55 \\
2038-04-09 23:37 $\pm$ 7.28 hr  & 2038-04-09 17:00 $\pm$ 7.15 hr  & 2038-04-10 05:12 $\pm$ 7.57 hr  & 13.08 $\pm$ 4.42 & 3.55 $\pm$ 3.25  & 0.19 $\pm$ 0.28 & 0.73 \\
2038-07-21 20:46 $\pm$ 22.81 hr & 2038-07-21 12:40 $\pm$ 22.09 hr & 2038-07-22 05:10 $\pm$ 21.28 hr & 17.53 $\pm$ 6.19 & 4.52 $\pm$ 4.57  & 0.03 $\pm$ 0.21 & 0.56 \\
2038-08-13 17:01 $\pm$ 7.60 hr  & 2038-08-13 12:07 $\pm$ 7.32 hr  & 2038-08-13 21:55 $\pm$ 7.52 hr  & 10.20 $\pm$ 3.40 & 3.02 $\pm$ 2.66  & 0.15 $\pm$ 0.22 & 0.72 \\
2038-11-24 20:06 $\pm$ 21.36 hr & 2038-11-24 08:09 $\pm$ 21.76 hr & 2038-11-25 05:21 $\pm$ 20.29 hr & 20.65 $\pm$ 7.65 & 4.27 $\pm$ 3.60  & 0.13 $\pm$ 0.22 & 0.66 \\
2038-12-17 11:15 $\pm$ 7.93 hr  & 2038-12-17 04:49 $\pm$ 7.67 hr  & 2038-12-17 17:19 $\pm$ 7.91 hr  & 12.93 $\pm$ 4.23 & 3.00 $\pm$ 2.46  & 0.27 $\pm$ 0.27 & 0.81 \\
2039-03-30 06:15 $\pm$ 22.87 hr & 2039-03-29 20:09 $\pm$ 22.26 hr & 2039-03-30 14:31 $\pm$ 21.57 hr & 18.97 $\pm$ 6.88 & 4.56 $\pm$ 4.48  & 0.09 $\pm$ 0.21 & 0.62 \\
2039-04-22 00:37 $\pm$ 7.70 hr  & 2039-04-21 18:27 $\pm$ 7.65 hr  & 2039-04-22 06:47 $\pm$ 7.91 hr  & 13.12 $\pm$ 4.46 & 3.14 $\pm$ 2.88  & 0.25 $\pm$ 0.26 & 0.80 \\
2039-08-03 00:53 $\pm$ 22.77 hr & 2039-08-02 16:33 $\pm$ 21.97 hr & 2039-08-03 08:15 $\pm$ 20.98 hr & 15.80 $\pm$ 5.84 & 4.42 $\pm$ 3.83  & 0.04 $\pm$ 0.20 & 0.59 \\
2039-08-25 18:54 $\pm$ 8.09 hr  & 2039-08-25 14:10 $\pm$ 7.73 hr  & 2039-08-25 23:44 $\pm$ 8.01 hr  & 9.83 $\pm$ 3.39  & 2.99 $\pm$ 2.42  & 0.17 $\pm$ 0.22 & 0.74 \\
2039-12-06 20:11 $\pm$ 21.46 hr & 2039-12-06 11:53 $\pm$ 21.62 hr & 2039-12-07 08:34 $\pm$ 20.25 hr & 19.80 $\pm$ 7.68 & 4.79 $\pm$ 3.63  & 0.10 $\pm$ 0.23 & 0.63 \\
2039-12-29 12:08 $\pm$ 8.38 hr  & 2039-12-29 06:20 $\pm$ 7.85 hr  & 2039-12-29 17:58 $\pm$ 8.24 hr  & 12.60 $\pm$ 4.13 & 3.26 $\pm$ 2.59  & 0.21 $\pm$ 0.28 & 0.77 \\
\enddata
\tablecomments{See Table \ref{tab:huya} for definitions and details of each column.}
\label{tab:xr}
\end{deluxetable*}
\end{longrotatetable}

\bibliography{all}{}
\bibliographystyle{aasjournalv7}



\end{document}